\documentclass[letter]{elsarticle}
\usepackage{amsmath,amssymb,amsfonts}
\usepackage{graphicx,subcaption,epsfig}
\usepackage{xcolor}
\usepackage{dcolumn}
\usepackage{multicol}
\usepackage{upquote,multirow,slashed}
\usepackage{appendix}
\usepackage{empheq}
\usepackage{lipsum}
\usepackage{booktabs,dsfont}
\usepackage{dcolumn}
\usepackage{xspace}
\usepackage{balance}

\definecolor{ForestGreen}{rgb}{0.15,0.70,0.15}
\usepackage[colorlinks=true]{hyperref}
\hypersetup{
 linktocpage=true,
 citecolor=ForestGreen,
 linkcolor=blue,
 urlcolor=blue}

\usepackage{geometry}
\usepackage{fancyhdr}
\fancypagestyle{firstpage}{%
	
	\lhead{}
	\rhead{\small CPTNP-2025-032}
}
\allowdisplaybreaks[1]

\newcommand{\NOdisplay}[1]{ }

\def\MSbar{\overline{\mathrm{MS}}}
\def\SigmaMass{m_{\sigma}}
\def\mMS{\overline{m}}
\def\mOS{m_{\mathrm{os}}}
\def\ZpOm{\mathrm{C}_{\overline{m}}}

\def\FSEbare{\Sigma_B(\slashed{p}, m_B, \hat{\mu})}
\def\RIMOM{\mathrm{RI/MOM}}
\def\RIpMOM{\mathrm{RI'/MOM}}

\newcommand{\loguos}{L_{\mathrm{os}}}
\newcommand{\logusm}{L_{\sigma}}
\newcommand{\logums}{L_{\mathrm{ms}}}

\begin{document}


\begin{frontmatter}

\title{
Trace-anomaly-subtracted $\sigma$-mass for Heavy Quarks and Matching from Lattice QCD 
}

\author[SDU]{Long Chen}
\ead{longchen@sdu.edu.cn}
\author[SDU]{Cong Zhao}

\address[SDU]{School of Physics, Shandong University, Jinan, Shandong 250100, China}

\begin{abstract}

We demonstrate that the leading IR-renormalon divergence in the perturbative pole mass of a massive quark resides entirely in the contribution from the trace anomaly of the energy-momentum tensor in QCD.
Consequently, the recently proposed trace-anomaly-subtracted $\sigma$-mass definition for heavy quarks is not only scheme- and scale-invariant, hence unique for each flavor, but also free from the leading IR-renormalon ambiguity. 
We further derive a formula connecting this $\sigma$-mass to the perturbative pole mass, solely in terms of the QCD $\beta$-function, quark-mass anomalous dimension $\gamma_m$ and a proper rewritten form of the pole-to-$\MSbar$ mass conversion factor. 
Utilizing this formula together with the ingredients available in the literature, we present the explicit five-loop result for the perturbative relationship between the $\sigma$-mass and the perturbative pole mass in QCD under the approximation of keeping only a single quark massive. 
Furthermore, we outline how $\sigma$-mass for a heavy quark can in principle be determined from lattice QCD through an intermediate regularization-independent renormalization and continuum perturbative matching, for which we analyze the matching factor through four loops with appealing features revealed. 
Given the aforementioned theoretical merits of this mass definition and the availability of high-precision conversion relations as well as the feasibility of direct extraction from lattice QCD, it is well suited for acting as a promising candidate reference scheme for comparing or combining quark masses of the same flavor obtained in different schemes.

\end{abstract}

\end{frontmatter}

\thispagestyle{firstpage}

Quark masses are fundamental parameters of the Standard Model of particle physics, 
and their precise values are of utter importance for various high-precision tests of this model. 
However, the color confinement inherent in Quantum Chromodynamics (QCD) makes the experimental extraction of quark masses highly non-trivial, as quarks are not observed as isolated free particles but are permanently confined within hadrons (with the exception of the top quark, which decays before hadronization).
Related to this, there is no theoretically unique way to define the ``physical'' mass of an isolated (heavy) quark. 
Nevertheless, within the framework of perturbative QCD, the pole mass of a massive quark can still be defined~\cite{Veltman:1992tm,Breckenridge:1994gs,Kronfeld:1998di,Gambino:1999ai}. 
However, it was discovered~\cite{Bigi:1994em,Beneke:1994sw} that the perturbative pole mass definition is plagued with a leading IR-renormalon issue, which prevents this definition from being meaningful to all orders in perturbation theory.

In the present work we are concerned with the leading infrared (IR) renormalon~\cite{tHooft:1977xjm,Parisi:1978bj,Parisi:1978az,David:1983gz,Mueller:1984vh} in the contribution from the trace anomaly of the energy-momentum tensor (EMT)~\cite{Crewther:1972kn,Chanowitz:1972vd,Chanowitz:1972da,Adler:1976zt,Collins:1976yq,Nielsen:1977sy} to the perturbative pole mass of a massive quark (which is itself defined to \textit{any} but only \textit{finite} order~\cite{Veltman:1992tm,Breckenridge:1994gs,Kronfeld:1998di,Gambino:1999ai} in perturbative QCD). 
And we aim to demonstrate that this contribution fully captures the leading IR-renormalon singularity observed in the perturbative pole mass definition~\cite{Bigi:1994em,Beneke:1994sw,Beneke:1994rs,Smith:1996xz}.
In addition, we outline how $\SigmaMass$ for a heavy quark can in principle be determined from lattice QCD through an intermediate regularization-independent (non-perturbative) renormalization and continuum perturbative matching at high orders, and reveal the appealing features of this approach.

\section{Leading IR-renormalon terms in the trace-anomaly contribution}

Regarding the trace-anomaly contribution to the perturbative pole mass in QCD  under the approximation of keeping only a single quark massive, one of us has derived the following relation~\cite{Chen:2025iul} at any loop order, 
\begin{equation}\label{eq:FFinsertion2FP} 
\langle p, s \big|\, 2\epsilon \big[-\frac{1}{4} F^a_{\mu\nu}\,F^{a\, \mu\nu}\big]_B \, \big| p, s \rangle \big|_{\mathrm{ampu.}} 
= \bar{u}(p, s) \Big( \hat{\mu} \frac{\partial\, \FSEbare}{\partial\, \hat{\mu} }\Big) \,u(p, s)\,,  
\end{equation} 
an identity between the dimensionally-regularized bare (unsubtracted) amputated matrix element of the EMT trace-anomaly operator $2\epsilon \big[-\frac{1}{4} F^a_{\mu\nu}\,F^{a\, \mu\nu}\big]_B$ over the on-shell massive quark state described by the Dirac spinor $u(p, s)$ with an on-shell momentum $\slashed{p} = \mOS$\footnote{With some abuse of notation, the shorthand equality $\slashed{p} = \mOS$ shall always be understood as implicitly applying to on-shell Dirac-spinors satisfying the on-shell equation of motion. 
} (and helicity $s$), and the bare self-energy function $\FSEbare$ in the Landau gauge. 
Equation~\eqref{eq:FFinsertion2FP} isolates the gluonic component of the complete bare EMT trace in full QCD (modulo equation-of-motion and BRST-exact operators).
Following an all-order diagrammatic proof of an exact identity between the pole mass and the connected on-shell matrix element for a massive quark~\cite{Chen:2025iul}, the remaining mass operator in the EMT trace gives the ``classical'' (Higgs-generated) mass contribution to the pole mass of a massive quark that is identified with its trace-anomaly-subtracted $\sigma$-mass. 
$\FSEbare$ is defined according to the usual parameterization of the full inverse propagator $\slashed{p} - m_B - \FSEbare$ of the massive quark; we refer to ref.~\cite{Chen:2025iul} for further technical details.
Although omitted from the compact notation, $\FSEbare$ also depends on the bare QCD coupling $\alpha_s^B$ and will be computed perturbatively as a power series in this parameter.
Having in mind the $\MSbar$ renormalization of $\alpha_s^B$ in Dimensional Regularization (DR) with spacetime $D = 4-2\epsilon$, we adopt the usual convention\footnote{Here it is unnecessary to pull out the conventional $e^{\epsilon \gamma_E} \big(4 \pi\big)^{-\epsilon}$ factor related to the particular choice of normalization convention for loop integration measures, which is irrelevant for the present discussion.} $\alpha_s^B \equiv \hat{\mu}^{2\epsilon}\, \hat{\alpha}_s^B$ for introducing a reduced mass-dimensionless bare coupling $\hat{\alpha}_s^B$ at the expense of introducing the auxiliary mass-dimensionful variable $\hat{\mu}$ in DR (which can be set conveniently, albeit not necessarily, to the same value as the actual renormalization or subtraction scale $\mu$).

It is very important to note that the logarithmic \textit{partial} derivative $\hat{\mu} \frac{\partial\, \FSEbare}{\partial\, \hat{\mu} }$ in the r.h.s.~of eq.~\eqref{eq:FFinsertion2FP} shall be understood as taken \textit{before} approaching the on-shell kinematic limit $\slashed{p} \rightarrow \mOS$. 
To be more specific, a slightly generalized form of eq.~\eqref{eq:FFinsertion2FP} can be stated in terms of the amputated Green correlation function with the insertion of the local operator $\mathcal{O}_{F}[\xi] \equiv \big[-\frac{1}{4}  F^{a}_{\mu \nu}\,F^{a\, \mu \nu} -\frac{1}{2 \xi} \big(\partial_{\mu} A^{\mu}_a \big)^2 \big]_B$ at zero momentum as follows: 
\begin{equation}\label{eq:FFinsertion2FP_Amp} 
\int \mathrm{d}^D x\, \mathrm{d}^D y \, e^{+i p \cdot (x-y)} \,
 \langle 0 \big|\, 2\epsilon\, \hat{\mathrm{T}} \{ \mathcal{O}_{F}[\xi]\, \psi_B(x)\, \bar{\psi}_B(y) \} \,\big| 0 \rangle_{\mathrm{ampu.}}  = \hat{\mu} \frac{\partial\, \FSEbare}{\partial\, \hat{\mu} } \,, 
\end{equation} 
which holds for an off-shell momentum $p$ in a generic covariant gauge-fixing condition parameterized by $\xi$.
At the on-shell limit $\slashed{p} \rightarrow \mOS$, eq.~\eqref{eq:FFinsertion2FP_Amp} reduces to eq.~\eqref{eq:FFinsertion2FP} in the Landau gauge corresponding to $\xi =0$.
The partial derivative w.r.t.\ $\hat{\mu}$ in the r.h.s.\ of eq.~\eqref{eq:FFinsertion2FP_Amp} shall be performed before $\slashed{p} \rightarrow \mOS$.
We shall take eq.~\eqref{eq:FFinsertion2FP} and/or eq.~\eqref{eq:FFinsertion2FP_Amp} as the starting point of the following discussion on the leading IR-renormalon terms in the quantum trace-anomaly contribution  $\hat{\mu} \frac{\partial\, \Sigma_B(\slashed{p}, m_B\,, \hat{\mu})}{ \partial\, \hat{\mu}} \big|_{\slashed{p} \rightarrow \mOS} $ so defined at the on-shell limit.
~\\

If we boldly \textit{assume} that one can exchange the operation ordering of taking the \textit{partial} derivative in $\hat{\mu}$ and approaching the on-shell momentum configuration $\slashed{p} = \mOS$,
we then obtain
\begin{equation}\label{eq:TAcontribution}
\hat{\mu} \frac{\partial\, \FSEbare}{ \partial\, \hat{\mu}} \big|_{\slashed{p} \rightarrow \mOS}
=  \hat{\mu} \frac{\partial\, \big(\mOS - m_B \big)}{ \partial\, \hat{\mu}}
= \hat{\mu} \frac{\partial\, \mOS(m_B, \hat{\mu})}{ \partial\, \hat{\mu}}\,,
\end{equation}
where we have employed the on-shell renormalization condition leading to
$\FSEbare \big|_{\slashed{p} = \mOS} = \mOS - m_B\,,$
and $\frac{\partial\, m_B}{ \partial\, \hat{\mu}} = 0$ holding owing to the bare mass being independent of the DR auxiliary scale $\hat{\mu}$.
The meaning of this partial derivative $\hat{\mu} \frac{\partial\, \mOS }{ \partial\, \hat{\mu}}$ shall be interpreted with care, as indicated by the last equality in \eqref{eq:TAcontribution} with the arguments specified explicitly in brackets---in particular, $\mOS(m_B,\hat{\mu})$ is henceforth shorthand for $\mOS(m_B,\hat{\alpha}_s^B,\hat{\mu})$, with the dimensionless bare coupling $\hat{\alpha}_s^B$ held fixed.
We note that this partial derivative differs conceptually from the total derivative w.r.t.\ the renormalization scale $\mu$ associated with the well-known renormalization-scale independence of the perturbative pole mass~\cite{Tarrach:1980up,Breckenridge:1994gs,Smith:1996xz,Kronfeld:1998di}, i.e.\ $\mathrm{d}\, \mOS / \mathrm{d}\, \mu = 0$.
The non-vanishing $\hat{\mu} \frac{\partial\, \mOS(m_B, \hat{\mu})}{ \partial\, \hat{\mu}} $ is itself among the possible reasons why taking the partial derivative in $\frac{\partial\, \FSEbare}{ \partial\, \hat{\mu}}$ may not, in general, naively commute with approaching the on-shell limit $\slashed{p} \rightarrow \mOS$.

We now justify the heuristic result in eq.~\eqref{eq:TAcontribution} with a more rigorous derivation.
To this end, let us start from the original unambiguous defining form for the on-shell renormalized operator matrix element introduced in eq.~\eqref{eq:FFinsertion2FP_Amp}, namely
\begin{equation}\label{eq:TAcontribution_def} 
\bar{u}(p, s)\, \mathrm{TA}_m \, u(p, s) \equiv \bar{u}(p, s)\, Z_{\psi}\, \Big( \hat{\mu} \frac{\partial\, \FSEbare}{\partial\, \hat{\mu} }\Big) \big|_{\slashed{p} \rightarrow \mOS}  \, u(p, s) \, 
\end{equation} 
where the partial derivative $\frac{\partial\,}{ \partial\, \hat{\mu}}$ is defined as taken \textit{before} approaching the on-shell limit $\slashed{p} \rightarrow \mOS$ (and the external on-shell wavefunction (re)normalization factor $Z_{\psi}$ is incorporated explicitly).
It is convenient at this moment to recall the on-shell renormalization condition in terms of the subtracted self-energy correction $\Sigma_R$ defined according to $ Z_{\psi}\,\big(\slashed{p} - m_B - \FSEbare \big) = \slashed{p} - \mOS - \Sigma_R $, which reads 
\begin{equation}\label{eq:osRC_SigR}
\Sigma_R\, \Big|_{\slashed{p} = \mOS} =0\,;\quad 
\frac{\partial\, \Sigma_R\,}{\partial\, \slashed{p}} \Big|_{\slashed{p} = \mOS} =0\,.
\end{equation} 
We can now proceed with the derivation in the following sequence:
\begin{align}  \label{eq:dMPodLogMU_origin}
\bar{u}(p, s)\, \mathrm{TA}_m \, u(p, s) & =\,
\bar{u}(p, s)\, 
Z_{\psi} 
\Big( 
\hat{\mu}\, \frac{ \partial\, \Big(\FSEbare + m_B  - \slashed{p} \Big)}{ \partial\, \hat{\mu}} \Big|_{\slashed{p} \rightarrow \mOS} 
\Big) 
 \, u(p, s) 
\nonumber\\
&=\,
\bar{u}(p, s)\, 
Z_{\psi} 
\Big( 
\hat{\mu}\, \frac{ \partial\,\Big( Z^{-1}_{\psi}\, \big(\mOS(m_B, \hat{\mu}) - \slashed{p} + \Sigma_R\, \big)  \Big)}{ \partial\, \hat{\mu}} \Big|_{\slashed{p} \rightarrow \mOS} 
\Big) 
 \, u(p, s) 
\nonumber\\
&=\,
\bar{u}(p, s)\, 
Z_{\psi} 
\Big( 
\hat{\mu}\, Z^{-1}_{\psi}\, \frac{ \partial\,\Big( \mOS(m_B, \hat{\mu}) - \slashed{p} + \Sigma_R\,  \Big)}{ \partial\, \hat{\mu}} \Big|_{\slashed{p} \rightarrow \mOS} 
\Big) 
\, u(p, s) 
\nonumber\\ 
&\,+\,
\bar{u}(p, s)\, 
Z_{\psi} 
\Big( 
\hat{\mu}\, \Big( \frac{ \partial\,Z^{-1}_{\psi}\,}{ \partial\, \hat{\mu}} \Big)\, \Big( \mOS - \slashed{p} + \Sigma_R\, \Big) \Big|_{\slashed{p} \rightarrow \mOS} 
\Big)
 \, u(p, s) 
\nonumber\\
&=\,
\bar{u}(p, s)\, \Big( 
\hat{\mu}\, \frac{ \partial\,\Big( \mOS(m_B, \hat{\mu}) - \slashed{p}  \Big)}{ \partial\, \hat{\mu}} \Big|_{\slashed{p} \rightarrow \mOS} 
\,+\,
\hat{\mu}\,\frac{ \partial\,\Sigma_R\, }{ \partial\, \hat{\mu}} \Big|_{\slashed{p} \rightarrow \mOS}  
\Big) \, u(p, s) 
\nonumber\\
&=\, 
\bar{u}(p, s)\, \hat{\mu}\, \frac{ \partial\,\mOS(m_B, \hat{\mu}) }{ \partial\, \hat{\mu}}  \, u(p, s) 
\,,
\end{align}
where the partial derivative w.r.t.~$\hat{\mu}$ is always taken \textit{before} approaching $\slashed{p} \rightarrow \mOS$. 
The vanishing of the second part in the third last equality of eq.~\eqref{eq:dMPodLogMU_origin} at $\slashed{p} = \mOS$ is due to the on-shell renormalization condition~\eqref{eq:osRC_SigR}, i.e.~the vanishing of the inverse renormalized propagator $\mOS - \slashed{p} + \Sigma_R $ at $\slashed{p} = \mOS$, and the existence of $\partial\,Z^{-1}_{\psi} / \partial\, \hat{\mu} $.
The vanishing of the second part in the second last equality of eq.~\eqref{eq:dMPodLogMU_origin} is due to the implication of~\eqref{eq:osRC_SigR}: $\Sigma_R$ has an asymptotic series expansion around the pole mass $\slashed{p} = \mOS$ that begins with the quadratic power-suppression factor $\mathcal{O}\big((\slashed{p}-\mOS)^2 \big)$. 
We thus manage to demonstrate in a rigorous manner that the heuristic result~\eqref{eq:TAcontribution} happens to be the correct result.
~\\

The task now is to properly interpret the meaning of this partial derivative $ \hat{\mu} \frac{\partial\, \mOS }{ \partial\, \hat{\mu}}$ in eq.~\eqref{eq:TAcontribution} and~\eqref{eq:dMPodLogMU_origin}. 
For the sake of investigating the IR-renormalon behavior~\cite{Bigi:1994em,Beneke:1994sw,Beneke:1994rs,Smith:1996xz} related to the pole-mass definition, we shall employ a parameterization in which this property is fully manifested.
To this end, we take the $\MSbar$-renormalized mass $\overline{m}$, which is commonly assumed in the literature to have no IR sensitivity and hence to be free from the (leading) IR-renormalon issue, since it is essentially the bare mass $m_B$ up to pure UV-poles in DR.
To be specific, the relationship between $\mOS$ and $\overline{m}$ can be established in dimensionally-regularized QCD via the following multiplicative relation with the same $m_B$:%
\begin{equation} \label{eq:massrelation2bare}
m_B = Z_{\overline{m}}(\alpha_s) \, \overline{m} = Z_{m}(\alpha_s,\, \mu/\mOS)\, \mOS 
\end{equation}%
where $\mu$ denotes the scale of $\MSbar$-renormalized QCD-coupling $\alpha_s(\mu)$, introduced via the usual $\MSbar$ renormalization of the reduced mass-dimensionless bare coupling: 
$\hat{\alpha}^B_s = \alpha_s(\mu)\, Z_{a_s} $ with $ \hat{\mu} = \mu $.
Consequently, we define the finite pole-to-$\MSbar$ mass conversion factor $\ZpOm$ by  
\begin{equation}\label{eq:Zpm_def}
\ZpOm\equiv  \ZpOm(\alpha_s,\, \mu/\overline{m}) = \frac{\mOS}{\overline{m}} = \frac{Z_{\overline{m}}}{ Z_{m}(\alpha_s,\, \mu/\mOS)} \Big|_{\epsilon \rightarrow 0}\,    
\end{equation}
where one shall insist on rewriting the explicit logarithmic mass-dependence in $\ZpOm$ using $\overline{m}$, rather than $\mOS$ (which would result in a different explicit $\mu$-dependence in the same  ratio).
This may be achieved in practice via an iterative application of $\ZpOm = \frac{\mOS}{\overline{m}}$ in the perturbative expansion of the r.h.s.~of eq.~\eqref{eq:Zpm_def}.

Let us now examine the partial derivative in eq.~\eqref{eq:dMPodLogMU_origin} in terms of the renormalized quantities.
In the $\MSbar$ scheme, $Z_{a_s}$ and $Z_{\overline{m}}$ have no explicit dependence on the scale $\hat{\mu}$. 
Hence, at fixed $(m_B,\hat{\alpha}_s^B)$, the relations $\hat{\alpha}_s^B=Z_{a_s}\alpha_s$ and $m_B=Z_{\overline{m}}\overline{m}$ imply that $\alpha_s$ and $\overline{m}$ are also held fixed under this particular partial derivative, namely $\frac{\partial\, Z_{\overline{m}}}{ \partial \mu} = 0 = \frac{\partial\, \overline{m}}{ \partial \mu} $, 
which shall not be confused with the RG total derivative. 
After taking $\epsilon\to0$ and setting $\hat{\mu}=\mu$, the operation is therefore transcribed into the derivative with respect to the explicit $\mu$ in $\mOS=\overline{m}\,\ZpOm(\alpha_s,\mu/\overline{m})$. 
Consequently, we may rewrite $ \hat{\mu} \frac{\partial\, \mOS }{ \partial\, \hat{\mu}}$ as:
\begin{align}\label{eq:MosPmuD}
\left.\mu \frac{\partial\, \mOS(m_B,\,\hat{\alpha}_s^B,\, \mu) }{ \partial\, \mu}\right|_{m_B,\,\hat{\alpha}_s^B}
&= \mu \frac{\partial\, \mOS(m_B = Z_{\overline{m}}\, \overline{m},\,\hat{\alpha}_s^B=Z_{a_s}\alpha_s,\, \mu) }{ \partial\, \mu}
 \,=\, \overline{m} \, \mu \frac{\partial\, \ZpOm(\alpha_s,\, \mu/\overline{m})}{ \partial\, \mu}\bigg|_{\overline{m},\,\alpha_s}\,,
\end{align}
where the partial derivative on the r.h.s.\ acts solely on the explicit $\mu$ in the ratio $\mu/\overline{m}$, with $\overline{m}$ and $\alpha_s$ held fixed.
Now comes the critical point.
We make use of a crucial well-established property of the leading IR-renormalon ($\mathrm{LIR}$) singularity observed in the perturbative pole mass definition of a massive quark, which can be formulated in terms of $\ZpOm$ defined in eq.~\eqref{eq:Zpm_def} as follows: 
the leading IR-renormalon terms in $\ZpOm$ --- expressed in terms of $\alpha_s(\mu)$ --- depend explicitly on $\mu$ only \textit{linearly}~\cite{Bigi:1994em,Beneke:1994sw,Beneke:1994rs,Beneke:1998ui}.
Consequently, we have 
\begin{eqnarray}\label{eq:MosPmuD_LIR}
\mu \frac{\partial\, \mOS(m_B,\, \mu) }{ \partial\, \mu} \Big|_{\mathrm{LIR}}
&=& \overline{m} \, \mu \frac{\partial\, \ZpOm(\alpha_s,\, \mu/\overline{m})}{ \partial\, \mu} \Big|_{\text{linear-$\mu$}}  \nonumber\\
&=& \overline{m} \, \ZpOm(\alpha_s,\, \mu/\overline{m}) \Big|_{\text{linear-$\mu$}} 
= \mOS(m_B,\, \mu) \Big|_{\mathrm{LIR}}\,.
\end{eqnarray}
We have thus succeeded in proving that the leading IR-renormalon terms in the trace-anomaly contribution to the perturbative pole mass $\mOS(m_B,\, \mu)$ of a massive quark, defined in eq.~\eqref{eq:TAcontribution} and~\eqref{eq:dMPodLogMU_origin}, are the same as those in $\mOS(m_B,\, \mu)$ itself.
~\\

An explicit formula can actually be derived for the above trace-anomaly contribution $\mu \frac{\partial\, \mOS(m_B,\, \mu) }{ \partial\, \mu} $ with the aid of the renormalization-group (RG) equation for the perturbative pole mass resulting from its scale-independence~\cite{Tarrach:1980up,Veltman:1992tm,Breckenridge:1994gs,Smith:1996xz,Kronfeld:1998di,Gambino:1999ai}, namely $\mu^2 \frac{\mathrm{d}\, \mOS}{ \mathrm{d}\, \mu^2 } = 0$, where $\mathrm{d}/\mathrm{d}\mu^2$ denotes the \textit{total} derivative that accounts for the implicit $\mu$ dependence via the running of $\alpha_s(\mu)$ and $\overline{m}(\mu)$.
More explicitly, the RG-equation reads 
\begin{align}\label{eq:ZmCallenSymanzik}
0 = \frac{\mu^2}{\ZpOm} \frac{\partial\,  \ZpOm\big(\alpha_s,\, \mu/\overline{m} \big)}{ \partial\, \mu^2 } 
+ \frac{\mu^2}{\ZpOm} \frac{\mathrm{d}\, \alpha_s(\mu) }{ \mathrm{d}\,  \mu^2}   \frac{\partial \ZpOm\big(\alpha_s\,, \mu/\overline{m}\big)}{\partial \alpha_s} 
+ \frac{\mu^2}{\ZpOm} \frac{\mathrm{d}\, \overline{m}(\mu) }{ \mathrm{d}\,  \mu^2}   \frac{\partial \ZpOm\big(\alpha_s\,, \mu/\overline{m}\big)}{\partial \overline{m}} 
+ \frac{\mu^2}{\overline{m}} \frac{\mathrm{d} \overline{m}(\mu)}{\mathrm{d} \mu^2}\,.     
\end{align} 
This can be turned into the following equation for the partial derivative of $\ZpOm$ w.r.t.~$\mu$:
\begin{eqnarray}
-\mu^2 \frac{\partial\,  \ln\big( \ZpOm\big(\alpha_s\,, \mu/\overline{m}\big) \big)}{ \partial\, \mu^2 }
&=& 
\gamma_m  \,+\, 
\beta\, \frac{\partial \ln\big( \ZpOm\big(\alpha_s\,, \mu/\overline{m}\big) \big)}{\partial \ln\big(\alpha_s\big)} 
+ \gamma_m \, \frac{\partial \ln\big( \ZpOm\big(\alpha_s\,, \mu/\overline{m}\big) \big)}{\partial \ln\big( \overline{m} \big)} \,,
\end{eqnarray}
where $\beta \equiv \mu^2\frac{\mathrm{d} \ln(\alpha_s)}{\mathrm{d} \mu^2}$ and $\gamma_m \equiv \mu^2\frac{\mathrm{d} \ln{\overline{m}(\mu)}}{\mathrm{d} \mu^2}$ denote, respectively, the anomalous dimensions of $\alpha_s$ and $\overline{m}$. 
Now owing to 
$$
\frac{\partial \ln\big( \ZpOm\big(\alpha_s\,, \mu/\overline{m}\big) \big)}{\partial \ln\big( \overline{m} \big)} = 
- 2 \mu^2 \frac{\partial\,  \ln\big( \ZpOm\big(\alpha_s\,, \mu/\overline{m}\big) \big)}{ \partial\, \mu^2 } \,,
$$
the RG-equation for the partial derivative of $\ZpOm$ w.r.t.~$\mu$ can be further reduced to the following form:   
\begin{eqnarray} \label{eq:ZmCallenSymanzik_red} 
- \mu^2 \frac{\partial\, \ln\big( \ZpOm\big(\alpha_s,\, \mu/\overline{m} \big) \big)}{ \partial\, \mu^2 } \big(1 - 2 \gamma_m \big) &=& \gamma_m  \,+\, 
\beta\, \frac{\partial \ln\big( \ZpOm\big(\alpha_s\,, \mu/\overline{m}\big) \big)}{\partial \ln(\alpha_s)}\,. 
\end{eqnarray}
We thus finally obtain the following explicit formula for the trace-anomaly contribution to the perturbative pole mass of a heavy quark: 
\begin{eqnarray}\label{eq:TAcont_formula}
\mu \, \frac{\partial\, \mOS}{ \partial\, \mu }  \Big|_{\alpha_s,\overline{m}}
&=& 
\mOS\, \Big( 2\mu^2 \frac{\partial\, \ln\big( \ZpOm\big(\alpha_s,\, \mu/\overline{m} \big) \big)}{ \partial\, \mu^2 } \Big) \nonumber\\
&=& 
\mOS\, \frac{-2 \, \gamma_m  \,-\, 
2\, \beta\, \frac{\partial \ln\big( \ZpOm\big(\alpha_s\,, \mu/\overline{m}\big) \big)}{\partial \ln(\alpha_s)} }{1 - 2 \gamma_m}  
\end{eqnarray}
in terms of the anomalous dimensions of $\alpha_s$ and $\overline{m}$ and the pole-to-$\MSbar$ conversion factor $\ZpOm$.
We emphasize once more that the partial derivatives of $\ZpOm\big(\alpha_s,\, \mu/\overline{m} \big) $ w.r.t.\ $\mu$ and/or $\alpha_s$ are defined by expressing $\ZpOm = \mOS/\overline{m}$ as a function of the independent variables $\alpha_s$ and $\mu/\overline{m}$ (rather than $\mu/\mOS$).
An appealing feature of \eqref{eq:TAcont_formula} is that, since $\beta$ has a perturbative expansion starting from $\mathcal{O}(\alpha_s^1)$, the perturbative result for $\mu \, \frac{\partial\, \mOS}{ \partial\, \mu } $ at $\mathcal{O}(\alpha_s^N)$ involves the perturbative expression of $\ZpOm $ only up to $\mathcal{O}(\alpha_s^{N-1})$, i.e.\ one loop order lower!

\section{Mass conversion formula for the trace-anomaly-subtracted $\SigmaMass$ of a heavy quark}

The mass-dimensional-analysis equation for the perturbative pole mass $\mOS(m_B,\, \hat{\mu})$ in QCD with only a single quark kept massive reads   
\begin{eqnarray}\label{eq:PoleBareMassMDA}
\mOS(m_B,\, \hat{\mu}) = \hat{\mu} \frac{\partial\, \mOS(m_B,\, \hat{\mu}) }{ \partial\, \hat{\mu}}
+  m_B \frac{\partial\, \mOS(m_B,\, \hat{\mu}) }{ \partial\, m_B}\,,
\end{eqnarray}
where both partial derivatives are taken at fixed $\hat{\alpha}_s^B$ and with the other displayed dimensionful variable held fixed.
The first term, $\hat{\mu} \frac{\partial\, \mOS(m_B,\, \hat{\mu}) }{ \partial\, \hat{\mu}}$, the pure trace-anomaly contribution, contains all the leading IR-renormalon terms in $\mOS(m_B,\, \hat{\mu})$, as we have just demonstrated above. 
On the other hand, one of us has proven~\cite{Chen:2025iul} that the forward on-shell matrix element of the EMT-trace operator over an elementary heavy quark state is \textit{identical} to its perturbative pole mass at any (finite) loop order in perturbative QCD where the incorporation of the trace-anomaly contribution is essential.
In the case of only one flavor of quark kept massive, it consists exactly of two pieces: the trace-anomaly contribution, as discussed in eq.~\eqref{eq:TAcontribution} and~\eqref{eq:MosPmuD}, and the remaining  ``classical'' fermion-mass operator part or the Higgs-generated mass contribution defined by $\SigmaMass = \mathrm{Z}_{\sigma} \, \mOS$ with 
$\mathrm{Z}_{\sigma} \equiv  \langle p, s  \big| \big[m \bar{\psi} \psi\big]_B  \, \big| p, s \rangle / \big(\bar{u}(p, s ) \, \mOS \, u(p, s) \big)$.
In view of eq.~\eqref{eq:PoleBareMassMDA}, we thus conclude that this trace-anomaly-subtracted $\SigmaMass$ for a heavy quark admits the following equivalent expression:
\begin{eqnarray}\label{eq:sigmass_altexp}
\SigmaMass &=& 
\mOS(m_B,\, \hat{\mu}) - \hat{\mu} \frac{\partial\, \mOS(m_B,\, \hat{\mu}) }{ \partial\, \hat{\mu}} \,=\, m_B\, \frac{\partial\, \mOS(m_B,\, \hat{\mu}) }{ \partial\, m_B}\,\nonumber\\
&=& 
\mOS - \overline{m} \, \mu \frac{\partial\, \ZpOm(\alpha_s,\, \mu/\overline{m})}{ \partial\, \mu} \,.
\end{eqnarray}
To arrive at the last equality, we have made use of properties of partial derivatives as similarly done in eq.~\eqref{eq:MosPmuD}.\footnote{We note in passing that the relation $\SigmaMass = m_B\, \frac{\partial\, \mOS(m_B,\, \hat{\mu}) }{ \partial\, m_B}$ complies formally with the so-called Feynman-Hellmann formula for composite quantum systems (we refer to refs.~\cite{Gasser:1979hf,Hoferichter:2025ubp} and the classical refs.~therein for more details).}

With the aid of the formula~\eqref{eq:TAcont_formula} for the trace-anomaly contribution, we thus end up with the following more explicit form for the ratio of the trace-anomaly-subtracted $\SigmaMass$ to $\mOS$, 
\begin{eqnarray}\label{eq:TASmass_formula}
\mathrm{Z}_{\sigma} \, = \, \frac{\SigmaMass}{\mOS} 
\, = \,
\frac{1 \,+\, 2\,
\beta\, \frac{\partial \ln\big( \ZpOm\big(\alpha_s\,, \mu/\overline{m}\big) \big)}{\partial \ln\big(\alpha_s\big)} }{1 - 2 \gamma_m }\,,
\end{eqnarray}
in terms of the anomalous dimensions of $\alpha_s$ and $\overline{m}$ and the pole-to-$\MSbar$ conversion factor $\ZpOm$.\footnote{We note that eq.~(2.8) given in ref.~\cite{Adler:1976zt} for the electron in QED cannot be applied here, owing to the different intermediate renormalization conditions employed.} 
We emphasize again that the partial derivative of $\ZpOm\big(\alpha_s,\, \mu/\overline{m} \big) $ w.r.t.\ $\alpha_s$ is defined by expressing $\ZpOm = \mOS/\overline{m}$ as a function of the independent variables $\alpha_s$ and $\mu/\overline{m}$ (rather than $\mu/\mOS$). 
The formula~\eqref{eq:TASmass_formula} derived with this identification also reproduces the known three-loop trace-anomaly contribution obtained by \textit{direct} diagrammatic calculation~\cite{Chen:2025iul}. 
We note an appealing feature of \eqref{eq:TASmass_formula}: owing to the leading perturbative term of $\beta$ being $\mathcal{O}(\alpha_s^1)$, the perturbative result for $\mathrm{Z}_{\sigma}$ at $\mathcal{O}(\alpha_s^N)$ involves the perturbative expression of $\ZpOm$ only up to $\mathcal{O}(\alpha_s^{N-1})$, i.e.~one loop-order less. 
In other words, the perturbative result for $\ZpOm$ at $N$-loop (i.e.~$\mathcal{O}(\alpha_s^N)$) is sufficient to derive the result for $\mathrm{Z}_{\sigma}$ at $N+1$-loop (i.e.~$\mathcal{O}(\alpha_s^{N+1})$), provided the knowledge of $\beta$ and $\gamma_m$ up to $\mathcal{O}(\alpha_s^{N+1})$.
With the relationship~\eqref{eq:TASmass_formula}, one can also readily obtain the conversion factor of  $\SigmaMass$ to the $\MSbar$ mass:
\begin{eqnarray}\label{eq:TASmass_over_MSbar}
\frac{\SigmaMass}{\mMS} =\mathrm{Z}_{\sigma}\, \ZpOm 
\, = \,
\frac{\ZpOm \,+\, 2\,
\beta\, \alpha_s\, \frac{\partial\,  \ZpOm\big(\alpha_s\,, \mu/\overline{m}\big)}{\partial\, \alpha_s} }{1 - 2 \gamma_m }\,.
\end{eqnarray}%
~\\

We note in passing that the formula~\eqref{eq:TASmass_formula} can also be employed to derive an explicit result for the on-shell quark-quark matrix element of the $\MSbar$-renormalized gluon-field strength squared $[F^{a}_{\mu\nu}\,F^{a\, \mu\nu}]_R$ that appears in the explicit all-order (operator-level) trace-anomaly formula~\cite{Adler:1976zt,Collins:1976yq,Nielsen:1977sy}, i.e.~$\Theta^{\mu}_{\,\mu} =  \frac{\beta}{2} [F^{a}_{\rho\sigma}\,F^{a\, \rho\sigma}]_R + (1 - 2 \gamma_m) [m \bar{\psi} \psi]_R $.
Explicitly, we have 
\begin{eqnarray}\label{eq:FFR_osme}
\langle p, s  \big| \big[ F^{a}_{\mu\nu}\,F^{a\, \mu\nu} \big]_R  \, \big| p, s \rangle  =  -4\, \frac{\partial \ln\big( \ZpOm\big(\alpha_s\,, \mu/\overline{m}\big) \big)}{\partial \ln\big(\alpha_s\big)} \, \bar{u}(p, s ) \, \mOS \, u(p, s)\,, 
\end{eqnarray}
where it is important to note that $[F^{a}_{\rho\sigma}\,F^{a\, \rho\sigma}]_R$ is purely $\MSbar$-renormalized~\cite{Nielsen:1975ph,Collins:1976yq,Nielsen:1977sy,Tarrach:1981bi,Spiridonov:1984br,Grinstein:1988wz}.

\section{Explicit perturbative result for $\SigmaMass/\mOS$ up to five loops in QCD}

The relationship between the perturbative pole mass and $\MSbar$ mass in QCD with a single massive quark has been derived up to three-loop order~\cite{Chetyrkin:1999qi} analytically~\cite{Melnikov:2000qh}, and to four-loop order~\cite{Marquard:2015qpa,Marquard:2016dcn}, albeit with a few four-loop non-logarithmic terms known only numerically (see ref.~\cite{Kataev:2019zfx} for the estimates of higher-order corrections).
Consequently, the formula~\eqref{eq:TASmass_formula} enables us to derive the relationship between the trace-anomaly-subtracted $\SigmaMass$ of a heavy quark and its perturbative pole mass $\mOS$ up to five-loop order, taking as inputs this four-loop pole-to-$\MSbar$ conversion factor\footnote{We take the numerical four-loop expression for the pole-to-$\MSbar$ conversion factor at $\mu = \mMS$ directly from refs.~\cite{Herren:2017osy,Kataev:2019zfx}, where a few non-logarithmic four-loop constant terms are currently known at per-mille-level accuracy. Consequently, the coefficients of the five-loop non-logarithmic terms in the result~\eqref{eq:TASoOS_5Loop} have a relative accuracy of about $0.2\%$ (with $n_l=5$).} and the state-of-the-art five-loop results for $\beta$~\cite{Baikov:2016tgj,Herzog:2017ohr,Luthe:2017ttg} and $\gamma_m$~\cite{Baikov:2014qja,Luthe:2016xec,Baikov:2017ujl}.
Specialized to the case of QCD with SU$(3)$ color group, we obtain the following numerical result evaluated at the scale $\mu = \mOS$:
\begin{align} \label{eq:TASoOS_5Loop}
\SigmaMass/\mOS &=\, 1
\,+\, \alpha_s\, \big(-0.636620 \big)
\,+\, \alpha_s^2\, \big(-1.11735 + 0.0731764 \,n_l \big)
\nonumber\\
&+\, \alpha_s^3\, \big(-4.98197 + 0.800055 \,n_l - 0.0206485 \,n_l^2 \big)
\nonumber\\
&+\, \alpha_s^4\, \big( -31.2996 + 6.70684 \,n_l - 0.405322 \,n_l^2 + 0.00658157 \,n_l^3 \big)
\nonumber\\
&+\, \alpha_s^5\, \big( -243.76(11) + 68.515(5)\, n_l  - 6.4963(2)\, n_l^2 + 0.240658\, n_l^3 -0.00295411\, n_l^4
\big) \,+\, \mathcal{O}(\alpha_s^6)\,,
\end{align}
where $n_l$ denotes the number of massless quark flavors included in the Lagrangian, and the parenthetical notations in the five-loop contribution list the errors inherited from the per-mille-level numerical uncertainties in the four-loop non-logarithmic piece of the pole-to-$\MSbar$ mass relation~\cite{Marquard:2015qpa,Marquard:2016dcn}.
The finite-order coefficients are expressed in the full-theory coupling $\alpha_s^{(n_f)}$ with $n_f=n_l+1$: the retained massive quark remains active, while the other $n_l$ active quarks are treated as massless in these coefficients. 
The full five-loop expression for $\SigmaMass/\mOS$ with exact numbers (apart from the aforementioned limitations)---too long to be presented in the text---is provided in an associated supplemental file, where the mass dependence in the logarithms has been consistently rewritten in terms of $\mOS$, i.e.~$\loguos \equiv \ln\big(\mu^2/\mOS^2 \big)$, but only \textit{after} applying the formula~\eqref{eq:TASmass_formula}.
The perturbative inverse of this relation is also derived and presented in the same supplemental file, where the mass dependence in the logarithms is consistently rewritten in terms of $\SigmaMass$, i.e.~$\logusm \equiv \ln\big(\mu^2/\SigmaMass^2 \big)$.
This latter result can be employed to conveniently transform an original perturbative expression for a physical observable involving $\mOS$ of heavy quarks into a function of $\SigmaMass$. 
One point is worthy of being pointed out explicitly here:
regarding the perturbative relation to the pole mass, the one loop, i.e.~$\mathcal{O}(\alpha_s)$,  correction to $\SigmaMass/\mOS$ is quite significant; 
once normalized by the size of $\mathcal{O}(\alpha_s)$ correction, the remaining relative higher-order perturbative corrections in \eqref{eq:TASoOS_5Loop} are generally smaller compared to, for instance, those in the case of the kinetic mass~\cite{Bigi:1994ga,Bigi:1996si,Czarnecki:1997sz,Fael:2020iea} up to three loops.

Furthermore, using the result~\eqref{eq:TASmass_over_MSbar} and the four-loop pole-to-$\MSbar$ conversion factor~\cite{Marquard:2015qpa,Marquard:2016dcn}, it is then straightforward to derive the relationship between the $\SigmaMass$ of a heavy quark and its $\MSbar$ mass up to four-loop order.
Again, rather than documenting the lengthy expression with exact numbers and logarithmic scale dependence, we list its numerical result evaluated at the scale $\mu = \mMS$, which reads 
\begin{align}\label{eq:TASoMS_4Loop}
\SigmaMass/\mMS &=\, 1 
\,+\, \alpha_s\, \big(-0.212207 \big) 
\,+\, \alpha _s^2\, \big(-0.0254365 - 0.0323361 n_l\big)
\nonumber\\
&\,+\, 
\alpha _s^3\, \big(0.268010 + 0.00994659 \, n_l + 0.000401805 \,  n_l^2\big) 
\nonumber\\
&\,+\,
\alpha _s^4\, \big(1.162(17) - 0.29899(37) \, n_l + 0.0240154 \, n_l^2 - 0.000380218 \,  n_l^3 \big)
\,+\, \mathcal{O}(\alpha_s^5)\,.
\end{align}
The full expression for $\SigmaMass/\mMS$ with exact numbers 
is again provided in the associated supplemental file, where the mass dependence in the logarithms is consistently expressed in terms of $\mMS$, i.e.~$\logums \equiv \ln\big(\mu^2/\mMS^2 \big)$.
Compared with the perturbative series~\eqref{eq:TASoOS_5Loop}, the increase in the perturbative coefficients in eq~\eqref{eq:TASoMS_4Loop} is significantly reduced, due to the absence of the leading IR-renormalon behavior in the latter relation.
It is also interesting to notice that the signs of the perturbative coefficients in the ratio $\SigmaMass/\mMS$ could get flipped depending on the value of $\mu^2/\mMS^2$.
To be specific, this happens for the $\mathcal{O}(\alpha_s)$ coefficient when $\mu$ is higher than about $1.4$ times $\mMS$, and similarly for the higher-order coefficients (depending on also the value of $n_l$). 
~\\

Ref.~\cite{Chen:2025iul} offered an all-order diagrammatic proof of the EMT-trace/pole-mass identity for elementary particles in perturbative gauge theories, directly computed the separate operator matrix elements and $\SigmaMass$ through three loops, together with numerical $t$-, $b$-, and $c$-quark results. 
The present work derives the compact all-order conversion formula~\eqref{eq:TASmass_formula} and the renormalized gluonic matrix element~\eqref{eq:FFR_osme}, extends $\SigmaMass/\mOS$ through five loops, and extends $\SigmaMass/\mMS$ through four loops;
furthermore, the conjecture made in ref.~\cite{Chen:2025iul} that $\SigmaMass$ for heavy quark is free of the leading IR renormalon is \textit{proved} in this work. 
The new $\mathcal{O}(\alpha_s^4)$ and $\mathcal{O}(\alpha_s^5)$ terms follow from the master formula and known pole--$\MSbar$ and RG ingredients.
A direct extension of the diagrammatic operator-insertion approach of ref.~\cite{Chen:2025iul} would require a new massive five-loop on-shell calculation, including explicit external-field renormalization; no such complete calculation is currently available. 
Equation~\eqref{eq:TASmass_formula} bypasses that calculation and reproduces the direct three-loop result.

In view of the small error of the current PDG-average value for the $t$-quark mass $m_t^{\mathrm{os}} = 172.56 \pm 0.31$~GeV~\cite{ParticleDataGroup:2024cfk}\footnote{We set aside, for the moment, the dispute within the high-energy physics community over interpreting this value as the perturbative pole mass of $t$-quark, related to the absence of free $t$-quarks in reality and the presence of the IR-renormalon singularity in its theoretical definition.} and the decent convergence of the truncated perturbative relation~\eqref{eq:TASoOS_5Loop} with $\alpha_s$ at the scale of $t$-quark mass, we update the $\sigma$-mass of $t$-quark to be 
$$\SigmaMass^t = 158.67 \pm 0.29~\text{GeV}$$ 
using the $\MSbar$-renormalized 6-flavor coupling $\alpha_s^{(6)}(m_t^{\mathrm{os}}) = 0.1076$ to be definite. 
This value follows from a standard application of the $\alpha_s$-running starting from $\alpha_s^{(5)}(m_z) = 0.1179$ at four-loop order with a conventional decoupling matching performed around $m_t$;
with \eqref{eq:TASoOS_5Loop}, alternative values for $\alpha_s^{(6)}(m_t^{\mathrm{os}})$ can be readily used as well. 
The error indicated in this result is mainly induced by the error of the input PDG-average $m_t^{\mathrm{os}}$, as the conventional QCD-scale uncertainty in the five-loop result~\eqref{eq:TASoOS_5Loop} is reduced to a negligible relative value of $\pm 3 \times 10^{-4}$. (The error associated with the input $\alpha_s$ value is not included separately.)
With the $\MSbar$-mass for $b$-quark $\overline{m}_b(\mu=\overline{m}_b) = 4.18^{\,+0.04}_{\,-0.03}$~GeV~\cite{ParticleDataGroup:2024cfk} and the 5-flavor coupling $\alpha_s^{(5)}(\overline{m}_b) = 0.2242$, determined using a four-loop running from $\alpha_s^{(5)}(m_{\mathrm{z}}) = 0.1179$, we then obtain, using the four-loop conversion relation~\eqref{eq:TASoMS_4Loop}, $$\SigmaMass^{b} = 3.97^{\,+0.04}_{\,-0.03}~\text{GeV}\,.$$ 
The conventional QCD-scale uncertainty in this result is about $0.4\%$ in relative terms, and the errors given in the upper/lower-script are dominated by those originating from the input values for $\overline{m}_b(\mu=\overline{m}_b)$.

\subsection*{The mass effect from internal massive quarks}
The mass effect of internal massive quarks enters only virtually and starts from two-loop corrections. 
With an appropriate choice of the low-energy effective QCD Lagrangian with the heavier quark fields \textit{integrated out} and lighter quark fields approximated as massless, the remaining quark-mass effects not explicitly accounted for in our single-massive treatment shall be relatively power-suppressed in quark-mass ratios, and start only from two-loop order, i.e.~$\mathcal{O}(\alpha_s^2)$. 
A rule-of-thumb estimate shows that the mass effect of the other massive quarks is well below one per mille for $\SigmaMass^t$, and below one percent for $\SigmaMass^b$ (mainly due to the internal $c$-quark loops).\footnote{The contributions from these purely virtual quark loops to the $\SigmaMass$ 
have a well-defined massless limit; consequently, the light-quark-loop contributions can be well approximated by their leading massless limit (namely neglecting all corrections suppressed by the squared light-quark masses).
On the other hand, the contribution from the mass term $m_h \bar{\psi}_h\psi_h$ of an \textit{internal} heavy-quark field $\psi_h$ in the infinitely large mass limit $m_h \rightarrow \infty$ can be treated conveniently using the heavy-quark effective field theories, see e.g.~refs.~\cite{Shifman:1978zn,Chetyrkin:1997un}. 
The leading effect can be readily read off based on the early discussions on the decoupling of internal heavy quarks in ref.~\cite{Shifman:1978zn}, while we defer a complete analysis to a future follow-up work. 
As stated above, in the present work we consider an effective QCD Lagrangian with just $n_f = n_l + 1$ active quarks where only one is kept massive. 
}

To further quantify these mass effects, we consider a concrete scenario of 5-flavor QCD (i.e.~without $t$-quark) where only the $b$-quark and $c$-quark are kept massive with mass ratio $\frac{\mOS^c}{\mOS^b} = \frac{3}{10}$. 
We carried out an explicit computation of $\SigmaMass^b$ and $\SigmaMass^c$ in terms of Feynman diagrams up to two loops with the on-shell renormalization constants taken from ref.~\cite{Fael:2020bgs}. 
We find that the internal $c$-quark mass effect amounts to shifting the ratio of $\SigmaMass^b$ to $\mOS^b$ \textit{downward} additionally by approximately $ 6\, \alpha^2_s / (4\pi)^2$ which evaluates to about $0.2\, \%$ around the $b$-quark mass scale. 
The ratio of $\SigmaMass^c$ to $\mOS^c$ is \textit{lowered} by about $24.8\, \alpha^2_s/(4\pi)^2$ by incorporating the internal $b$-quark mass effect, which evaluates to about $1.7\, \%$ around the $c$-quark mass scale. 
Thus the numerical impacts of the masses of internal quarks for the $\SigmaMass$ for the $b$-quark and $c$-quark are found to be less than the percent-level errors induced by the residual QCD-scale uncertainties  remaining at three loops~\cite{Chen:2025iul}, but comparable to those at four loops (still less than the error induced by those in the input parameters). 
Therefore, we do not explicitly include these effects in the present analysis.
On the other hand, this treatment itself can also be conveniently \textit{defined} as a particular renormalization scheme for the heavy-quark mass of a particular flavor, where all quarks other than the particular heavy-quark flavor in question are considered in the chiral massless limit.

\section{Extracting $\SigmaMass$ for heavy quarks from lattice QCD via $\RIMOM$}

In high-energy particle physics, there exist several distinct definitions of quark masses other than the perturbative pole mass and $\MSbar$-mass, which include, e.g.~, the kinetic mass~\cite{Bigi:1994ga,Bigi:1996si,Czarnecki:1997sz,Fael:2020iea}, the potential-subtracted mass~\cite{Beneke:1998rk,Beneke:2005hg}, the 1S-mass~\cite{Hoang:1998ng}, the MSR-mass~\cite{Hoang:2008yj,Hoang:2017suc}, the (minimal) renormalon-subtracted mass~\cite{Pineda:2001zq,Ayala:2014yxa,Komijani:2017vep}, the $\RIMOM$ mass~\cite{Martinelli:1994ty} and RI/(m)SMOM mass~\cite{Aoki:2007xm,Sturm:2009kb,Boyle:2016wis,Chen:2025seb} (commonly used in lattice QCD). 
Even for the same quark flavor, the mass values obtained under different definitions generally differ and are typically extracted from the most suitable experimental observables or lattice-QCD quantities, chosen according to sensitivity and precision. 
To directly compare or combine (e.g., take averages of) numerical results obtained under these different mass definitions, it is necessary to first convert all of them into a single, \textit{common reference mass} scheme.
For this purpose, the $\MSbar$-mass scheme has so-far been widely adopted as the universal common reference scheme for quark masses. 
This choice is motivated by the scheme's freedom from infrared renormalon ambiguities, its short-distance nature with \textit{mass-independent} renormalization, and its role as the standard framework in perturbative QCD calculations (especially for high-energy scattering processes), thereby enabling consistent comparisons and combinations across different theoretical and lattice determinations. 
To this end, the perturbative relations among different quark masses are needed (see, e.g.~the recent comprehensive review~\cite{Beneke:2021lkq} and the compilations in refs.~\cite{Chetyrkin:2000yt,Herren:2017osy}); 
in particular, the relation between $\SigmaMass$ and these masses can be readily derived up to three or even four loops with the explicit results~\eqref{eq:TASoOS_5Loop} and~\eqref{eq:TASoMS_4Loop}.
However, the original definition of the $\MSbar$-subtraction~\cite{Bardeen:1978yd} scheme is intimately connected to dimensional regularization~\cite{tHooft:1972tcz,Bollini:1972ui} and cannot be \textit{directly} used with lattice regularization. 
In addition, the meaning of its associated normalization or subtraction scale is not physically transparent (which leads to the well-known ambiguities when considering the decoupling of heavy particles).

The regularization-independent momentum-subtraction scheme~\cite{Martinelli:1994ty}($\RIMOM$), as well as its variants and the like~\cite{Aoki:2007xm,Sturm:2009kb,Boyle:2016wis,Chen:2025seb}, is a special instance of the wide class of momentum-subtraction schemes --- whose precursors date back to the early days of quantum field theories --- that is free of the above shortcomings and can be implemented in lattice-QCD calculations.
With the aid of an intermediate regularization-independent scheme, the $\overline{\mathrm{MS}}$-subtraction and the resulting $\MSbar$-mass~\cite{Bardeen:1978yd} can be effectively implemented in any other regularization method provided the bridging $\RIMOM$ scheme, or any of its variants, can be implemented. 
In a similar spirit, a strategy for directly extracting $\SigmaMass$ for heavy quarks 
from lattice-QCD calculations can be formulated, in contrast to its apparent or superficial dependence on the concept of the on-shell pole mass (which is tricky to be properly defined in the non-perturbartive treatment).
The renormalization of the (short-distance) lattice bare mass $m_B^{\mathrm{lat}}$ for a heavy quark to its $\SigmaMass$ can be denoted and parameterized as   
\begin{equation}\label{eq:ZsmLattice}
\SigmaMass = Z_{\SigmaMass}^{lat}\,  m_B^{\mathrm{lat}} =  \mathrm{C}_{\mathrm{RI}\rightarrow\SigmaMass} \, Z_{m_{\mathrm{RI}}}^{lat} \,  m_B^{\mathrm{lat}} 
\end{equation}
where all UV divergences in the $\SigmaMass$-renormalization constant $Z_{\SigmaMass}^{lat} = \mathrm{C}_{\mathrm{RI}\rightarrow\SigmaMass} \, Z_{m_{\mathrm{RI}}}^{lat}$ are regularized via the intermediate $\RIMOM$ renormalization constant $Z_{m_{\mathrm{RI}}}^{lat}$ computed in lattice QCD. 
(The dependence of $m_B^{\mathrm{lat}}$ on the UV-regulating lattice spacing is suppressed in the  notation; this dependence is canceled against that in $Z_{m_{\mathrm{RI}}}^{lat}$ and replaced by momentum-subtraction scale dependence.) 
The finite renormalization or matching coefficient $\mathrm{C}_{\mathrm{RI}\rightarrow\SigmaMass} = \frac{\SigmaMass}{m_{\mathrm{RI}}}$, which effectively relates $\SigmaMass$ to $\RIMOM$-mass $m_{\mathrm{RI}} \equiv Z_{m_{\mathrm{RI}}}^{lat} \,  m_B^{\mathrm{lat}}$, can be conveniently determined in perturbative QCD with dimensional regularization.  
Consequently, once the lattice bare mass $m_B^{\mathrm{lat}}$ for a heavy quark is determined from lattice-QCD calculations, the renormalization constant $Z_{\SigmaMass}^{lat}$ can be applied to it to directly extract the $\SigmaMass$ for this heavy quark.

We have determined the perturbative result for the matching factor (in closed form) up to three-loop order as   
\begin{align}\label{eq:CmRI2SM}
\mathrm{C}_{\mathrm{RI}\rightarrow\SigmaMass}(\mu_s)        
&= 
1 \,+ \, a_s\, \Big(
4 L_m - 8
\Big) 
\nonumber \\ 
&  \,+\, 
a_s^2\, \Big(
n_l \, \left(\frac{16 L_{\mu }}{3}-\frac{8 L_{\mu } L_m}{3}-\frac{4 L_m^2}{3}-\frac{4 L_m}{9}-\frac{8 \pi ^2}{9}+\frac{122}{9}\right)
\nonumber \\ &\,+\, 
\left(48 \zeta _3-\frac{248 L_{\mu }}{3}+\frac{124 L_{\mu } L_m}{3}+\frac{86 L_m^2}{3}+\frac{50 L_m}{9}+\frac{16 \pi ^2}{3}-\frac{5111}{18}+\frac{16}{9} \pi ^2 \log (2)\right)
\Big) 
\nonumber \\ 
&  \,+\, 
a_s^3\, \Big(
n_l^2 \, \Big(\frac{64 \zeta _3}{9}-\frac{32 L_{\mu }^2}{9}+\frac{32 \pi ^2 L_{\mu }}{27}-\frac{488 L_{\mu }}{27}+\frac{16}{9} L_{\mu } L_m^2+\frac{16}{9} L_{\mu }^2 L_m+\frac{16 L_{\mu } L_m}{27}
\nonumber \\ &\,+\, 
\frac{16 L_m^3}{27}+\frac{8 L_m^2}{27}+\frac{32 \pi ^2 L_m}{27}-\frac{536 L_m}{81}+\frac{16 \pi ^2}{81}-\frac{1892}{81}\Big)
\nonumber \\ &\,+\, 
n_l \, \Big(-\frac{8096 \zeta _3}{27}-64 \zeta _3 L_{\mu }+\frac{992 L_{\mu }^2}{9}-\frac{688 \pi ^2 L_{\mu }}{27}+\frac{20522 L_{\mu }}{27}-\frac{16}{27} \pi ^2 \log (16) L_{\mu }-\frac{448 \zeta _3 L_m}{9}
\nonumber \\ &
-\frac{592}{9} L_{\mu } L_m^2-\frac{496}{9} L_{\mu }^2 L_m-\frac{1816 L_{\mu } L_m}{27}-\frac{640 L_m^3}{27}-\frac{1244 L_m^2}{27}-\frac{784 \pi ^2 L_m}{27}+\frac{16400 L_m}{81} -\frac{16}{27} \pi ^2 \log (16) L_m 
\nonumber \\ &\,+\,
\frac{512 \text{Li}_4\left(\frac{1}{2}\right)}{27}+\frac{560 \pi ^4}{243}-\frac{3416 \pi ^2}{81}+\frac{105260}{81}+\frac{64 \log ^4(2)}{81}+\frac{64}{81} \pi ^2 \log (2) \log (4)-\frac{320}{81} \pi ^2 \log (2)\Big)
\nonumber \\ &\,+\,  
\Big(-\frac{5756 \pi ^2 \zeta _3}{27}+\frac{413351 \zeta _3}{108}+\frac{6920 \zeta _5}{27}+992 \zeta _3 L_{\mu }-\frac{7688 L_{\mu }^2}{9}+\frac{992 \pi ^2 L_{\mu }}{9}-\frac{177737 L_{\mu }}{27}
\nonumber \\ &\,+\,
\frac{992}{27} \pi ^2 \log (2) L_{\mu }+\frac{752 \zeta _3 L_m}{9}
+\frac{5332}{9} L_{\mu } L_m^2+\frac{3844}{9} L_{\mu }^2 L_m+\frac{12748 L_{\mu } L_m}{27}+\frac{6364 L_m^3}{27}+\frac{13166 L_m^2}{27}
\nonumber \\ &\,+\,
\frac{1184 \pi ^2 L_m}{9}-\frac{192746 L_m}{81}+\frac{1184}{27} \pi ^2 \log (2) L_m-\frac{14080 \text{Li}_4\left(\frac{1}{2}\right)}{27}-\frac{1318 \pi ^4}{243}
\nonumber \\ &\,+\,
+\frac{1003282 \pi ^2}{1215}-\frac{7730803}{648}-\frac{1760 \log ^4(2)}{81}-\frac{1408}{81} \pi ^2 \log ^2(2)-\frac{25120}{81} \pi ^2 \log (2)\Big)
\Big) 
\nonumber \\ &\,+\,  
\mathcal{O}(a_s^4)\,, 
\end{align}
where a handy variable $a_s \equiv \frac{\alpha_s }{4 \, \pi}$ is introduced.
In principle, the full result for $\mathrm{C}_{\mathrm{RI}\rightarrow\SigmaMass}(\mu_s)$ depends only on the $\RIMOM$ subtraction scale $\mu_s$ accompanied by $\SigmaMass$ in terms of the logarithm $L_m \equiv \ln(\mu_s^2/\SigmaMass^2)$, and its residual (net) dependence on the $\alpha_s$-renormalization scale-$\mu$ in $L_{\mu} \equiv \ln(\mu^2/\mu_s^2)$ is merely a consequence of the fixed-order perturbative truncation.  
The dependence on $\mu_s$ is solely induced by that of $m_{\mathrm{RI}}(\mu_s)$, governed by the renormalization-group equation of $m_{\mathrm{RI}}(\mu_s)$~\cite{Chetyrkin:1999pq}, with the following logarithmic derivative (or anomalous dimension) up to four loops: 
\begin{align}\label{eq:CmRI2SMamd} 
\frac{\mu_s^2}{\mathrm{C}_{\mathrm{RI}\rightarrow\SigmaMass}}\, \frac{\mathrm{d}\,\mathrm{C}_{\mathrm{RI}\rightarrow\SigmaMass }} {\mathrm{d}\,\mu_s^2}  &= 
a_s\, \big( 4 \big)  
\,+\,a_s^2\, \Big(
n_l \, \left(-\frac{8 L_{\mu }}{3}-\frac{52}{9}\right)
\,+\, \left(\frac{124 L_{\mu }}{3}+\frac{1082}{9}\right)
\Big) 
\nonumber \\ 
&  \,+\, a_s^3\, \Big(
n_l^2 \, \left(\frac{16 L_{\mu }^2}{9}+\frac{208 L_{\mu }}{27}+\frac{928}{81}\right)
\,+\,
n_l \, \left(\frac{128 \zeta _3}{9}-\frac{496 L_{\mu }^2}{9}-\frac{8920 L_{\mu }}{27}-\frac{53302}{81}\right)
\nonumber \\ &\,+\, 
 \left(-\frac{9904 \zeta _3}{9}+\frac{3844 L_{\mu }^2}{9}+\frac{76732 L_{\mu }}{27}+\frac{510367}{81}\right)
\Big) 
\nonumber \\ &  
\,+\, a_s^4\,
\Big(
n_l^3 \, \left(-\frac{32 L_{\mu }^3}{27}-\frac{208 L_{\mu }^2}{27}-\frac{1856 L_{\mu }}{81}-\frac{18832}{729}\right)
\nonumber \\ &\,+\, 
n_l^2 \, \left(-\frac{6416 \zeta _3}{27}-\frac{256 \zeta _3 L_{\mu }}{9}+\frac{496 L_{\mu }^3}{9}+\frac{4352 L_{\mu }^2}{9}+\frac{49726 L_{\mu }}{27}+\frac{217372}{81}\right)
\nonumber \\ &\,+\, 
n_l \, \left(\frac{601463 \zeta _3}{54}+\frac{4160 \zeta _5}{3}+\frac{23776 \zeta _3 L_{\mu }}{9}-\frac{7688 L_{\mu }^3}{9}-\frac{78520 L_{\mu }^2}{9}-\frac{1030034 L_{\mu }}{27}-\frac{62527001}{972}\right)
\nonumber \\ &\,+\,  
\left(-\frac{13772281 \zeta _3}{108}+3440 \zeta _5-\frac{307024 \zeta _3 L_{\mu }}{9}+\frac{119164 L_{\mu }^3}{27}+\frac{1289042 L_{\mu }^2}{27}+\frac{17935423 L_{\mu }}{81}+\frac{2315030437}{5832}\right)
\Big) 
\nonumber \\ &\,+\,  
\mathcal{O}(a_s^5)\,.  
\end{align}
To facilitate the practical matching procedure with $\RIMOM$ inputs, the appearance of $\SigmaMass$ in $ \ln(\mu_s^2/\SigmaMass^2)$ can be rewritten in terms of $m_{\mathrm{RI}}(\mu_s)$ followed by a consistent perturbative re-expansion and truncation (which will lead to a minor numerical difference). 
Alternatively, this equation~\eqref{eq:CmRI2SM} can be numerically solved in an iterative manner, provided that it converges to a stable result after a few iterations.
Furthermore, although \eqref{eq:CmRI2SM} is independent of the $\alpha_s$-renormalization scale-$\mu$ (as $\SigmaMass$ is scale-independent), due to the availability of only a fixed-order truncated form, different choices of $\mu$ do matter a bit leading to perturbative truncation errors (albeit to a lesser extent at higher orders), which can be conveniently, but need not be, set equal to $\mu_s$.
With ingredients available in refs.~\cite{Chetyrkin:1999pq,Gracey:2022vjc,Bi:2023pnf,Bali:2020isn}, we provide the following explicit numerical results for the so-reformulated $\mathrm{C}_{\mathrm{RI}\rightarrow\SigmaMass} (\mu_s)$ evaluated at a benchmark point $\mu = \mu_s = 2\, m_{\mathrm{RI}}$ and $n_f = n_l + 1 = 4$: 
\begin{align}\label{eq:CmRI2SM4Lnum}
\mathrm{C}_{\mathrm{RI}\rightarrow\SigmaMass}\Big|_{\mu_s = 2\, m_{\mathrm{RI}}} 
 =  1 -0.19535 \,\alpha_s-0.46989 \,\alpha_s^2-0.90492 \,\alpha_s^3-1.835(18) \, \alpha_s^4 \,+\,\mathcal{O}(\alpha_s^5)  
\end{align}
which exhibits better apparent convergence than its $\MSbar$-counterpart~\cite{Chetyrkin:1999pq,Gracey:2022vjc,Bi:2023pnf,Bali:2020isn}.
The results up to four loops with full dependence on $\ln(\mu_s^2/m_{\mathrm{RI}}^2)$ and $n_l$ can be found in the attached electronic file.
As the $\RIMOM$-subtraction scale $\mu_s$ increases, the improvement in the apparent convergence of the truncated perturbative series~\eqref{eq:CmRI2SM4Lnum} is actually faster than that of its $\MSbar$-counterpart: 
in addition to the common benefit from the reduction of $\alpha_s$ value at higher scales, the perturbative coefficients themselves also become smaller.

One crucial advantage with this approach is that both $\mathrm{C}_{\mathrm{RI}\rightarrow\SigmaMass} (\mu_s)$ and its logarithmic derivative~\eqref{eq:CmRI2SMamd} in the $\RIMOM$ subtraction scale $\mu_s$ do \textit{not} rely on and are, in principle, independent of the $\MSbar$-renormalized perturbative $\alpha_s$ at a particular $\MSbar$ scale $\mu$ (which cannot be too small). 
Consequently, provided some accurate boundary values for both $\mathrm{C}_{\mathrm{RI}\rightarrow\SigmaMass} (\mu_s)$ and its logarithmic derivative~\eqref{eq:CmRI2SMamd} chosen at an initial point $\mu^{o}_s$ where the available truncated series in $\alpha_s(\mu)$ can be reliably applied, one can then solve the differential equation~\eqref{eq:CmRI2SMamd} in $\mu_s$ to determine the $\mathrm{C}_{\mathrm{RI}\rightarrow\SigmaMass} (\mu_s)$ in a region possibly far from its initial perturbative domain --- without involving $\alpha_s$ 
there --- if no singularities are encountered along the way.
Therefore, by exploiting this unique property, the involvement of the perturbative $\alpha_s(\mu)$ may be reduced and confined to merely the initial boundary values where this concept does reliably work. 
Based on the above discussions, an optimal determination of $\mathrm{C}_{\mathrm{RI}\rightarrow\SigmaMass}$ at a generic $\RIMOM$-subtraction scale $\mu_s$ can be envisaged as follows: 
take as the initial value for $\mathrm{C}_{\mathrm{RI}\rightarrow\SigmaMass}$ its perturbative value evaluated at an initial boundary $\mu^{o}_s$ where the truncated series~\eqref{eq:CmRI2SM} exhibits a good apparent convergence with a suitably chosen $\mu$, for instance $\mu \sim \mu_s \sim 2\, m_{\mathrm{RI}}$ around $2$~or $3$~GeV for the $c$-quark case\footnote{It is not necessary to take $\mu$ equal to $\mu_s$; for a given $\mu_s$ (and $m_{\mathrm{RI}}(\mu_s)$) one could roam around to search for an optimal scale $\mu$ based on a chosen criterion for good apparent convergence.}, and then derive its values at other $\mu_s$ of interest in the calculations by solving the differential equation~\eqref{eq:CmRI2SMamd}. 
The difference caused by whether the $\mathcal{O}(\alpha_s^4)$ term in \eqref{eq:CmRI2SM4Lnum} is included or not in the determination of the initial values could be taken as a conservative estimate of the perturbative truncation error in the so-derived $\mathrm{C}_{\mathrm{RI}\rightarrow\SigmaMass}$ used at three-loop accuracy. 
An analysis similar to the above can also be performed for the variant $\RIpMOM$~\cite{Martinelli:1994ty,Chetyrkin:1999pq}, 
where we have observed a slightly better perturbative convergence behavior.\footnote{The explicit results up to four loops can be found in the attachment, which we refrain from documenting in the text.}
Both this and the fact that the above matching factor is closer to 1 imply that it shall be advantageous, at least in terms of numerical uncertainties, to directly extract the numerical values for the $\SigmaMass$ for $c$-quark and $b$-quark from lattice QCD, rather than going through the intermediate scale-dependent $\MSbar$ mass at $2\,$GeV. 
It shall be interesting to compare the numerical values for the $\SigmaMass$ for heavy quarks (such as $c$-quark and $b$-quark) to the averaged $\sigma$-term contribution per heavy quark to the masses of the heavy-flavored hadrons in future studies.\footnote{We thank Y.B.~Yang for an enlightening discussion on this.} 
Given the theoretical merits of this mass definition --- unique for each quark flavor and free of the leading IR-renormalon --- and the availability of high-precision conversion relations as well as the feasibility of a direct accurate extraction from lattice-QCD calculations via~\eqref{eq:ZsmLattice} and \eqref{eq:CmRI2SM}, we conclude that it is well suited for acting as a promising unique common reference scheme for comparing or combining quark masses of the same flavor obtained under different definitions.

\section{Conclusion} 

We have discovered that the leading IR-renormalon divergence in the perturbative pole mass of a massive quark~\cite{Bigi:1994em,Beneke:1994sw,Beneke:1994rs,Smith:1996xz} resides entirely in the contribution from the trace anomaly of EMT in QCD.
Consequently, the trace-anomaly-subtracted $\sigma$-mass definition proposed in ref.~\cite{Chen:2025iul} for heavy quarks, which is scheme/scale-independent and further proved to be free from the leading IR-renormalon issue in this note, nicely combines the merits of both the perturbative pole-mass and $\MSbar$-mass definitions, while elegantly circumvents their respective unappealing and undesirable features.
Given the theoretical merits of this mass definition --- unique for each quark flavor and free of the leading IR-renormalon --- and the availability of high-precision conversion relations as well as a systematically improvable lattice-QCD matching strategy~\eqref{eq:ZsmLattice} and \eqref{eq:CmRI2SM}, we conclude that it is well suited for acting as a unique common reference scheme for comparing or combining quark masses of the same flavor obtained under different definitions. 

\section*{Acknowledgements}

The work of L.~C. was supported by the Natural Science Foundation of China under contract No.~12205171, No.~12235008, No.~12321005, and grants from the Department of Science and Technology of Shandong province tsqn202312052 and 2024HWYQ-005.
The authors gratefully acknowledge the valuable discussions and insights provided by the members of the China Collaboration of Precision Testing and New Physics. 

\bibliographystyle{utphysM}
\balance
\biboptions{sort&compress}

\bibliography{TASmass} 


\end{document}